\documentclass[conference]{IEEEtran}

\usepackage{amsfonts}
\usepackage{amssymb}
\usepackage{stfloats}
\usepackage{cite}
\usepackage{graphicx}
\usepackage{psfrag}
\usepackage{subfigure}
\usepackage{amsmath}
\usepackage{array}
\usepackage{algorithm}
\usepackage{algpseudocode}
\usepackage[affil-it]{authblk}
\usepackage[english]{babel}
\usepackage{blindtext}
\usepackage{url}
\usepackage{epstopdf}
\usepackage{mathrsfs}

\title{A Deep Learning Based Resource Allocation Scheme in  Vehicular Communication Systems}

\author{Mimi Chen$^{\dagger}$, Jiajun Chen$^{\dagger}$, Xiaojing Chen$^{\dagger}$, Shunqing Zhang$^{\dagger}$ and Shugong Xu$^{\dagger}$\\
$^{\dagger}$ Shanghai Institute for Advanced Communication and Data Science, \\
Key laboratory of Specialty Fiber Optics and Optical Access Networks, \\
Joint International Research Laboratory of Specialty Fiber Optics and Advanced Communication, \\
Shanghai University, Shanghai, 200444, China\\
Email:\{ChenMi, JiajunChen, jodiechen, shunqing, shugong\}@shu.edu.cn}

\begin{document}

\maketitle

\begin{abstract}
In vehicular communications, intracell interference and the stringent latency requirement are challenging issues. In this paper, a joint spectrum reuse and power allocation problem is formulated for hybrid vehicle-to-vehicle (V2V) and vehicle-to-infrastructure (V2I) communications. Recognizing the high capacity and low-latency requirements for V2I and V2V links, respectively, we aim to maximize the weighted sum of the capacities and latency requirement. By decomposing the original problem into a classification subproblem and a regression subproblem, a convolutional neural network (CNN) based approach is developed to obtain real-time decisions on spectrum reuse and power allocation. Numerical results further demonstrate that the proposed CNN can achieve similar performance as the Exhaustive method, while needs only 3.62\% of its CPU runtime.

\end{abstract}


\section{Introduction}
Recently, vehicle-to-vehicle (V2V) communications have attracted increasing attention for its potential to improve road safety and traffic efficiency, and enable delay-sensitive vehicular applications, where communications happen only between neighboring vehicles \cite{ashraf2016dynamic}. To improve the spectral efficiency in vehicular communications, V2V links have been designed to share the same radio resources of vehicle-to-infrastructure (V2I) uplinks. Intracell interference control between V2V and V2I links hence becomes an important issue in V2V communications.
To cope with the interference in vehicular communication systems, resource allocation strategies have been proposed in \cite{liang2017resource,sun2016radio,abbas2018hybrid}. The throughput of V2I links was maximized with a minimum quality-of-service (QoS) guarantee by performing spectrum sharing and power allocation for V2V and V2I links \cite{liang2017resource,sun2016radio}. 

On the other hand, to support delay-sensitive and high reliable information exchange, especially in the context of Ultra Reliable Low Latency Communications (URLLC) in future fifth-generation (5G) mobile communication system, latency is a particularly crucial requirement for V2V connections. To this end, many works have focused on the low-latency vehicular communications \cite{sun2014d2d,yu2017radio,abbas2018novel}. 
However, the works \cite{sun2014d2d,yu2017radio,abbas2018novel} neglected the QoS requirements of V2I links, thus are inapplicable to the scenario where V2V and V2I coexist. Consider a hybrid V2V and V2I communication scenario, \cite{mei2018latency} maximized the information rate of V2I links utilizing the Lagrange dual decomposition and binary search, with a considerable complexity. 

Most of the previous works \cite{liang2017resource,sun2016radio,abbas2018hybrid,sun2014d2d,yu2017radio,abbas2018novel,mei2018latency} derived the resource allocation schemes as the solutions of optimization problems, where iterative algorithms are applied. In iterative schemes, a large number of iterations need to be carried out before convergence is achieved. The high
computational cost prevents implementing these algorithms in real-time for practical uses.
As a key technique in artificial intelligence, deep learning has been widely used in image processing and voice processing \cite{krizhevsky2012imagenet,ren2015faster}. It has also been recently developed to solve traditional problems in wireless communications. Deep neural networks (DNNs) can be used to solve complex nonlinear non-convex problems without building complicated mathematical models. For example, \cite{Sun2017Learning} proposed a DNN-based algorithm to approximate a traditional iterative algorithm (i.e., WMMSE \cite{shi2011iteratively}) for real-time wireless resource management. 

In this paper, we propose a convolutional neural network (CNN) based resource allocation approach for hybrid V2I and V2V communications. The main contributions of this work can be summarized as follows.
\begin{itemize}
    \item  Considering a hybrid V2I and V2V communication scenario, we formulate the
resource allocation task as a joint spectrum reuse and power allocation problem. Recognizing the stringent latency requirement for V2V links, we maximize the weighted sum of the capacities and latency requirement for vehicular communications.
    \item Different from \cite{Sun2017Learning}, \cite{wang2018machine}, which either solve a regression problem or a classification problem for resource
allocation by deep learning, the proposed CNN, for the first time, decomposes the original problem of vehicular communications into a classification subproblem and a regression subproblem, to infer the optimal decisions on joint spectrum reuse and power allocation.
    \item Extensive numerical experiments are conducted to demonstrate that the proposed CNN can achieve similar
performance as the Exhaustive method, while substantially reduce the computational time. The low complexity makes the proposed approach well suited for high-speed mobile scenes in  vehicular communication.
\end{itemize}


The rest of this paper is organized as follows. Section~\ref{sect:System Model} describes the system model. The proposed neural network architecture is developed in Section III. Simulation results are provided in Section IV, followed by the conclusion in Section~\ref{sect:conc}.

\section{System Model} \label{sect:System Model}

\begin{figure}
\centering
\includegraphics[width=3.4in]{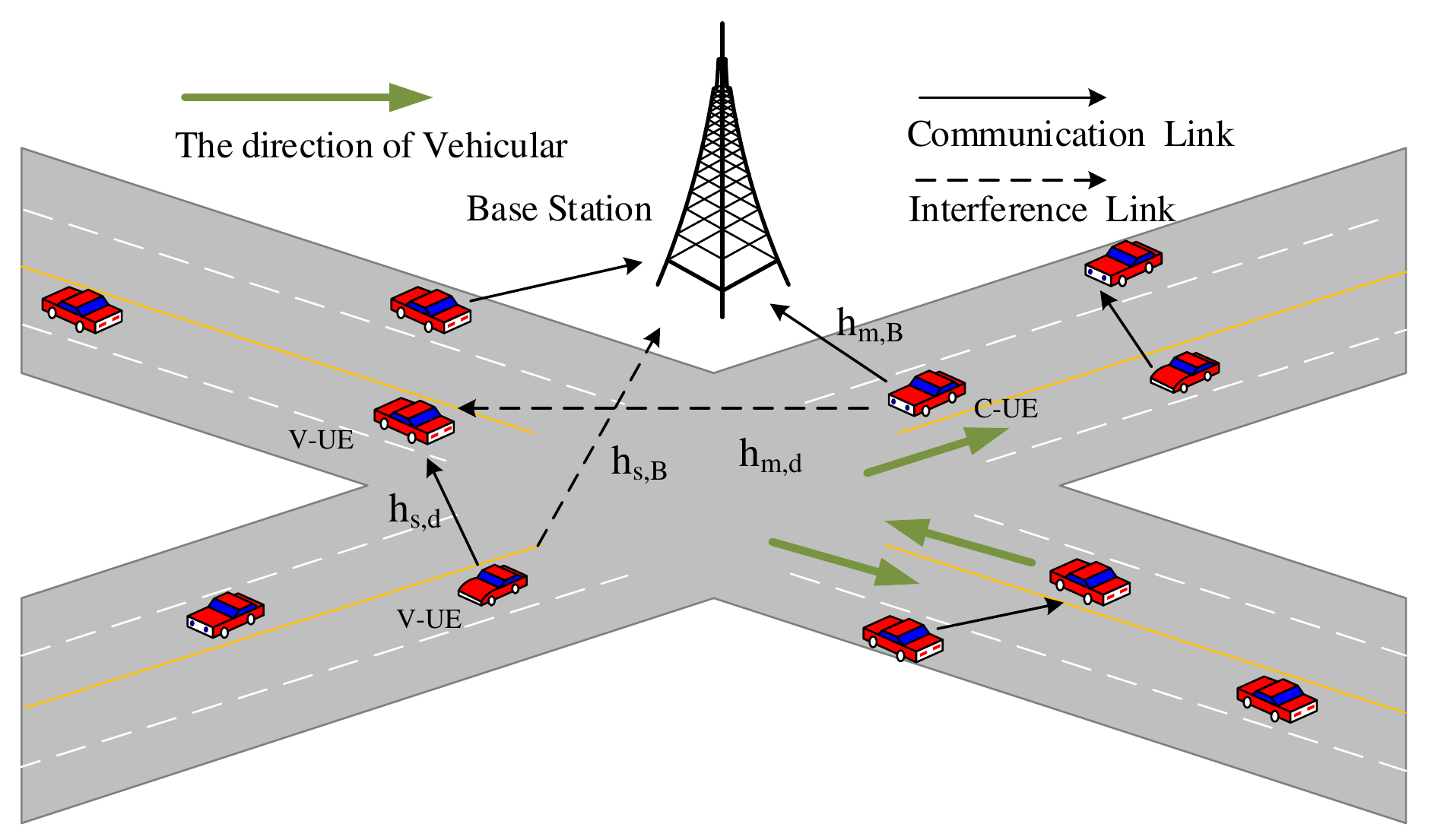}
\caption{A hybrid V2I and V2V communication scenario.}
\label{scenery}
\end{figure}

Consider a hybrid V2I and V2V transmission scenario as shown in Fig.~\ref{scenery}, where the vehicles are in the coverage of a single BS. $M$ vehicles are communicating with the BS through V2I links, denoted as cellular user equipment (C-UEs), while $N$ pairs of vehicles are exchanging data directly through V2V links, denoted as vehicular user equipment (V-UEs). For illustration purpose, we denote $\mathcal{M}:=\{1, 2, \ldots, M\}$, $\mathcal{S}:=\{1, 2, \ldots, N\}$, and $\mathcal{D}:=\{1, 2, \ldots, N\}$ as the vehicle sets for C-UEs, V-UE transmitters, and V-UE receivers, respectively. 
Moreover, we assume that each vehicle plays one of the following roles: i) C-UEs; ii) V-UE transmitters; iii) V-UE receivers; or iv) idle vehicles. 

To improve the communication reliability and the spectrum utilization, we assume that the uplink spectrum resources allocated orthogonally to the C-UEs can be reused by the V-UEs. To avoid introducing severe interference to cellular links, we assume that the spectrum resources of a C-UE can only be reused by one V-UE, and one V-UE can only access the spectrum of one C-UE. 
The channel power gain, h$_{m,B}$, between the $m$th C-UE and the BS can be expressed as
\begin{eqnarray}
h_{m,B} = g_{m,B}\alpha_{m,B},
\end{eqnarray}
where $g_{m,B}$ is the small-scale fast fading power component, assumed to be exponentially distributed with unit mean;
$\alpha_{m,B}$ is the large-scale fading power component consisting of pathloss and shadowing. The channel power gain $h_{s,d}$ between the $s$th V-UE transmitter and the $d$th V-UE receiver, the interference channel power gain $h_{s,B}$ between the $s$th V-UE transmitter and the BS, and the interference channel power gain $h_{m,d}$ between the $m$th C-UE and the $d$th V-UE receiver are similarly defined.


Let $\gamma_{m}^{c}$ and $\gamma_{d}^{v}$ denote the received signal-to-interference-plus-noise-ratios (SINRs) at the BS and the $d$th V-UE receiver, respectively. They can be given as
\begin{eqnarray}\label{sinr1}
\gamma_{m}^{c}=\frac{P_{m}^{c}h_{m,B}}{N_{0}+\sum\limits_{s\in{\mathcal{S}},d\in{\mathcal{D}}}\rho_{m,s}P_{s}^{v}h_{s,B}},
\end{eqnarray}
\begin{eqnarray}\label{sinr2}
\gamma_{d}^{v}=\frac{P_{s}^{v}h_{s,d}}{N_{0}+\sum\limits_{m\in{\mathcal{M}}}\rho_{m,s}P_{m}^{c}h_{m,d}},
\end{eqnarray}
where $P_{m}^{c}$ and $P_{s}^{v}$ denote the transmit power of the $m$th C-UE and the $s$th V-UE transmitter, respectively; $N_{0}$ is the noise power, and $\rho_{m,s}$ is the status of the spectrum resource reuses.  $\sum_{s\in{\mathcal{S}},d\in{\mathcal{D}}}\rho_{m,s}P_{s}^{v}h_{s,B}$ in \eqref{sinr1} denotes the received interference at the BS from the $s$th V-UE transmitter, and $\sum_{m\in{\mathcal{M}}}\rho_{m,s}P_{m}^{c}h_{m,d}$ in \eqref{sinr2} denotes the received interference at the $d$th V-UE receiver from the $m$th C-UE. Here, $\rho_{m,s} = 1$ means that the $s$th V-UE transmitter reuses the spectrum resource allocated to the $m$th C-UE, and $\rho_{m,s} = 0$ otherwise. 

Let $C_{m}$ and $C_{s}$ denote the ergodic capacities of the $m$th C-UE and the $s$th V-UE transmitter, which are calculated by the long-term average over the fast fading~\cite{Liang2017Resource}, as given by
\begin{eqnarray}
C_{m} = \mathbb{E}[\log_{2}(1 + \gamma_{m}^{c})|{g_{m,B}}, m\in \mathcal{M}],
\end{eqnarray}
\begin{eqnarray}
C_{s} = \mathbb{E}[\log_{2}(1 + \gamma_{d}^{v})|{g_{s,d}}, s\in \mathcal{S},d\in \mathcal{D}],
\end{eqnarray}
where $\mathbb{E}$ is the expectation taken over the fast fading distribution.

V2V links are often used to transmit urgent information to avoid collisions between vehicles. Hence, the communication latency is considered as one of the most important requirements for V2V links. In this paper, we denote $B$ and $L$ as the average packet size and the tolerable transmission latency, respectively. The target transmit rate of V-UEs is then given by $R = B/L$. To describe the latency requirement of the V2V links, we pick the smallest ergodic capacity $C_{ss}$ among V-UEs, and calculate the probability when the capacity of this V-UE is larger than $R$ over fast fading. Then the latency requirement can be expressed as
\begin{eqnarray}
\xi = P_L\{C_{ss}(k)\geq{R}\}, ~~k=1, 2, \ldots
\end{eqnarray}

Additionally, we set a minimum capacity requirement for the C-UEs to guarantee a minimum predetermined QoS.
Our objective is to maximize the weighted sum of the ergodic capacities of the V2I and V2V links, and the latency requirement of the V2V links, by making optimal decisions on the spectrum reuse $\rho_{m,s}$ and
power allocation $\{P_{m}^{c},P_{s}^{v}\}$. The resource allocation problem can be formulated as

\begin{subequations}
\begin{align}
\max\limits_{\{\rho_{m,s},P_m^c,P_s^v\}}&\sum\limits_{m\in{\mathcal{M}}}C_{m}+\omega_{1}\sum\limits_{s\in{\mathcal{S}},d\in{\mathcal{D}}}C_{s}+\omega_{2}\xi \label{objective}\\
\text{s.t.}~~&~~\omega_{1},\omega_{2} > 0,\\
&~~\log(1+\gamma^c_{m})\geq{r^c_0},\quad\forall{m}\in\mathcal {M}, \label{7c}\\
&~~~0\leq{P^c_{m}}\leq{P^c_{\max}},\quad~\forall{m}\in\mathcal {M},\label{7d}\\
&~~~0\leq{P^v_{s}}\leq{P^v_{\max}},\quad~~\forall{n}\in\mathcal {S},\label{7e}\\
&\sum\limits_{{s}\in\mathcal{S},d\in{\mathcal{D}}}\rho_{m,s}\leq1,\quad\quad\forall{m}\in\mathcal {M},\label{7f}\\
&~~\sum\limits_{{m}\in\mathcal{M}}\rho_{m,s}\leq1,\quad\quad~ \forall{s}\in\mathcal {S},\label{7g}\\
&~~\rho_{m,s}\in\left\{0,1\right\},\quad\quad~~~\forall{m}\in\mathcal {M}, {s}\in\mathcal {S}
\end{align}
\end{subequations}
where $\omega_{1}$ and $\omega_{2}$ are the weight factors, $r^c_0$ is the minimum required capacity for each V2I link. $P^c_{\max}$ and $P^v_{\max}$ are the maximum transmit powers of the V2I links and V2V links, respectively. \eqref{7c} represents the minimum capacity constraint to ensure the QoS of the V2I links. Constraints \eqref{7d} and \eqref{7e} make sure that the transmit powers of the C-UEs and V-UEs cannot exceed the maximum transmit powers. And Constraints \eqref{7f} and \eqref{7g} represent that the spectrum of one V2I link can only be reused by one V-UE, and one V-UE can only access the spectrum of a single V2I link.

The proposed formulation above not only realizes maximizing the weighted sum of the ergodic capacities of the V2I and V2V links in the vehicle communication system, but also guarantees the low-latency requirement of the V2V links. Unfortunately, this is a highly nonlinear non-convex optimization problem, which is in general very difficult to solve. Introducing the emerging deep learning technique, here we propose a CNN based approach to solve this problem.

\section{Deep Learning for Resource Allocation}\label{sect:dnn}
In this section, we first introduce the data generation phase, and then describe the proposed network structure and the chosen loss functions.

\subsection{Data Generation}

The setting of the hybrid V2I and V2V transmission network can be found in Section~IV.
Given pregenerated channel gains $\boldsymbol{h}:=\{h_{m,B}, h_{m,d}, h_{s,d}, h_{s,B}, \forall m, d, s\}$ and predetermined parameters $P^c_{\max}$, $P^v_{\max}$ and $r_0^c$, we generate the corresponding spectrum resource reuse state $\{\rho_{m,s}, \forall{m, s}\}$ and the allocated powers $\{P^c_m, P^v_s, \forall{m,s}\}$ for each channel realization by running an exhaustive method. The Exhaustive method iteratively calculates and compares the objective in~\eqref{objective} for all possible schemes and chooses one of the scheme that maximizes the objective as the optimal solution. By doing so, the Exhaustive method sets a benchmark for the proposed CNN-based approach with a high computational cost. By repeating the above process for multiple times, we generate the entire training data set $\{\boldsymbol{h}, \rho_{m,s}, P^c_m, P^v_s, \forall{m, d, s}\}$.

Different from existing works which either solve a regression
problem \cite{Sun2017Learning}, or a classification problem \cite{krizhevsky2012imagenet} for resource allocation by deep learning, our proposed CNN architecture decomposes the original problem (7) into a multi-label classification subproblem for spectrum reuse selection, and a regression subproblem for power allocation, and then outputs the joint optimal decisions.

Collect $\{\rho_{m,s}, \forall{m, s}\}$ in a $M\times N$ matrix, $\boldsymbol{A}$.   
The different values of $\boldsymbol{A}$, each of which associated with one solution of spectrum reuse, are classified into different classes. Each class is indexed by a one-hot encoded vector as its label, say $\rho_j$. Here, $j$ is the index number of all classes.



\subsection{Proposed Network Architecture}
\begin{figure*}[h]
\centering
\includegraphics[width=7in]{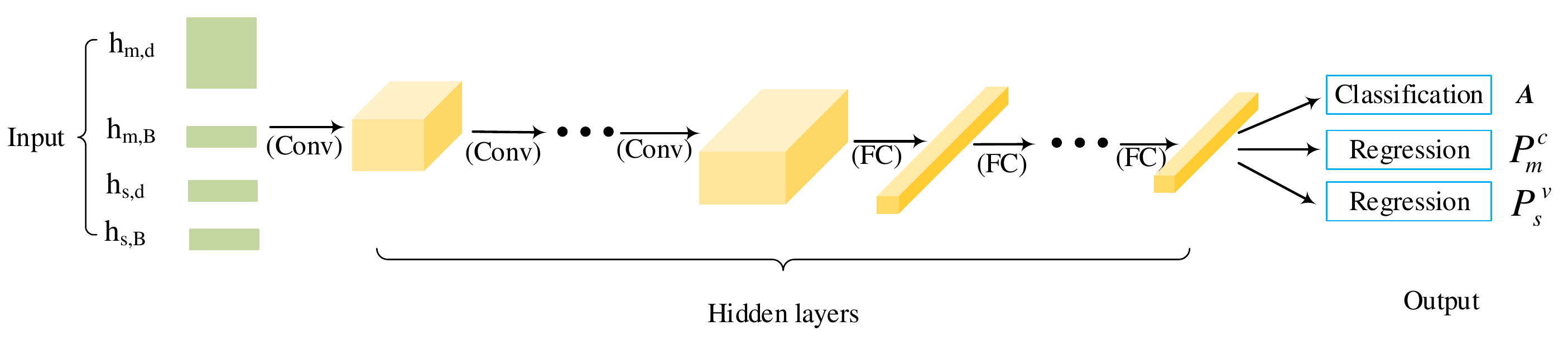}
\caption{The proposed CNN architecture that consists of one input layer, multiple hidden layers, and one output layer. }
\label{The Network Structure}
\end{figure*}

As shown in Fig.~\ref{The Network Structure}, the proposed CNN architecture consists of one input layer, multiple hidden layers, and one output layer.
\begin{itemize}
\item{\em \textbf{Input layer:}} The input data of the CNN are the channel gains $\boldsymbol{h}=\{h_{m,B}, h_{m,d}, h_{s,d}, h_{s,B}, \forall m, d, s\}$.
\item{\em \textbf{Hidden layers:}} The hidden layers are composed of three convolution layers and three Fully Connected (FC) layers with the activation function, \emph{Rectified Linear Unit} (RelU). In this paper, we consider a CNN as the training network for the reason that a CNN always outperforms other neural networks (e.g., a FC DNN) in feature extraction and highly accurate classification. It can exploit the spatial features in channel gains and reduce the number of weights compared to a FC DNN, such that real-time decisions on resource allocation can be made for practical uses.
\item{\em \textbf{Output layer:}} Three outputs are to be obtained from this layer. The first output is the class index of matrix $\boldsymbol{A}$ indicating spectrum reuse, which is the solution of a multi-label classification subproblem. We select \emph{softmax} as the activation function. The other two outputs are the transmit powers of the V2I and V2V links, $P^c_m$ and $P^v_s$, respectively. They are the outputs of a regression subproblem with the activation function, RelU.
\end{itemize}

We use the training data set to optimize the weights of the CNN. The CNN is trained to regenerate the decisions on spectrum reuse and power allocation derived from the Exhaustive method, given channel gains $\boldsymbol{h}$. 

\subsection{Loss Function}

As the proposed CNN aims to solve different subproblems (i.e., classification and regression), different loss functions are considered adapting to different features of the subproblems.

\begin{itemize}
\item{\em \textbf{Mean Squared Error (MSE):}} The loss function, MSE, is a reflection of the model's fitting degree to the training data. It can be described as
    \begin{eqnarray}
    L_{reg} = \frac{\parallel{\Tilde{\boldsymbol{P}}-\boldsymbol{P}}\parallel^2 }{K}= \frac{\sum_{i}{(\Tilde{P_i}-P_i)^2}}{K},
    \end{eqnarray}
    
    where $K$ is the number of batch size, $\Tilde{\boldsymbol{P}}$ is the predicted power vector of C-UEs (or V-UEs) output by the CNN, and $\boldsymbol{P}$ is the allocated power in the training set. The intuitive meaning of this loss function $L_{reg}$ is quite clear: the greater the Euclidean distance between the predicted value $\Tilde{\boldsymbol{P}}$ and the true value $\boldsymbol{P}$, the greater the loss, and vice versa.
\item{\em \textbf{Categorical Crossentropy:}} This is the loss function for the multi-label classification subproblem, as given by
    \begin{eqnarray}
    L_{cls} = \sum\limits_{j}\rho_{j}{\log{(\Tilde{\rho}_{j})}},
    \end{eqnarray}
    
where $\Tilde{\rho}_{j}$ is the predicted class index of spectrum reuse output by the CNN, and $\rho_{j}$ is the target class index of spectrum reuse.
\end{itemize}

Therefore, the CNN is trained to minimize the following total loss function:
\begin{eqnarray}
L = L_{cls} + \alpha{L_{reg}}(P^c_m) + \beta{L_{reg}}(P^v_s),
\end{eqnarray}
where $\alpha$ and $\beta$ are the weights for the loss functions of the transmit powers $P^c_m$ and $P^v_s$, respectively.

\section{Simulation Results} \label{sect:sim}
\subsection{Simulation Parameter}

\begin{figure}[h]
\centering
\includegraphics[width=3.4in]{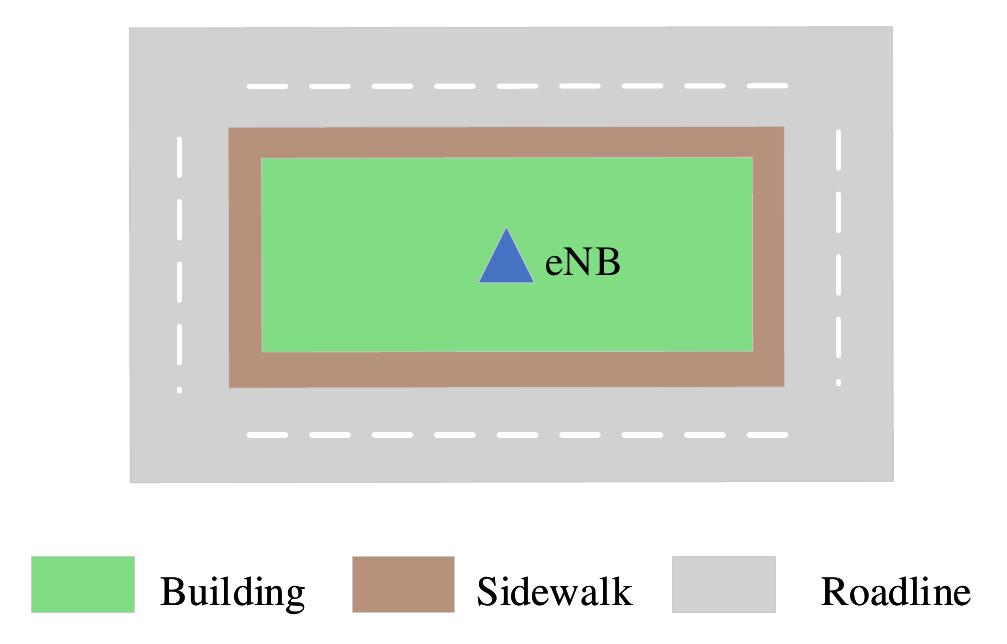}
\caption{Road configuration for urban vehicular communication.}
\label{urban}
\end{figure}

In the data generation phase, we consider a single cell urban scenario with the V2V case based on the Manhattan grid layout \cite{3gpp.36.885}. As shown in Fig.~\ref{urban}, the building size is 413~m $\times$ 30~m with 3~m reserved for sidewalk along the building. It is assumed that there are 2 lanes in each direction and the lane width is set as 3.5 m. The vehicles are dropped on the roads randomly following the spatial Poisson process. Table~\ref{channel mode} gives the channel models for the V2I and V2V links, and Table~\ref{Simulation Parameters} lists the rest parameters used in our simulation.

\begin{table}[h]
\newcommand{\tabincell}[2]{\begin{tabular}{@{}#1@{}}#2\end{tabular}}
\centering
\caption{Channel Models for V2I and V2V links}
\begin{tabular}{|c|c|c|}\hline
Parameter                  &V2I Link            &V2V Link\\\hline
Pathloss model              & \tabincell{c}{128.1 + 37.6$\log_{10}$\textsl{d},\\ \textsl{d} in km}  
                           & \tabincell{c}{WINNER + B1\cite{meinila2009winner} \\Manhattan grid layout}\\\hline
Shadowing distribution     & Log-normal         & Log-normal\\\hline
Shadowing std deviation   & 8 dB                & 3 dB\\\hline
Fast fading                & Rayleigh fading    & Rayleigh fading\\\hline
\end{tabular}
\label{channel mode}
\end{table}

\begin{table}[h]
\newcommand{\tabincell}[2]{\begin{tabular}{@{}#1@{}}#2\end{tabular}}
\centering
\caption{Simulation Parameters}
\begin{tabular}{|c|c|c|}
\hline
Parameter                     &value\\\hline
Carrier frequency             & 2 GHZ\\\hline
Bandwidth                     & 10 MHZ\\\hline
BS antenna height             & 25 m\\\hline
Vehicle antenna height        & 1.5 m\\\hline
Absolute vehicle speed        & 30 km/h\\\hline
Number of V-UE pairs $N$             & 5\\\hline
Number of C-UEs $M$              & 5\\\hline
Minimum capacity of C-UEs $r^c_0$            & 0.5 bps/HZ\\\hline
Maximum transmit power of C-UEs $P_{\max}^c$       & 23 dBm\\\hline
Maximum transmit power of V-UEs $P_{\max}^v$       & 23 dBm\\\hline
minimum transmit power of C-UEs $P_{\min}^c$       & 10 dBm\\\hline
minimum transmit power of V-UEs $P_{\min}^v$       & 10 dBm\\\hline
Noise power $N_0$        & -114 dBm\\\hline
Average packet size           & 6400 bits\\\hline
Maximum latency           & 100 ms\\\hline
\end{tabular}
\label{Simulation Parameters}
\end{table}


To better evaluate the performance of the proposed CNN-based
approach, we compare it with five other schemes: 1) Benchmark by using the Exhaustive method, which serves as an ideal reference; 2) DNN by using a FC DNN, with the parameters specified in Table~\ref{Training and Testing Parameters}; 3) RandomPower by randomly generating the power allocation following a uniform distribution; 4)
MaxPower by allocating the maximum transmit power for vehicles; and 5) MinPower by allocating the minimum transmit power for vehicles.
The latter three schemes serve as heuristic baselines.

\begin{table}[h]
\newcommand{\tabincell}[2]{\begin{tabular}{@{}#1@{}}#2\end{tabular}}
\centering
\caption{Training and Testing Parameters}
\begin{tabular}{|c|c|c|c|}
\hline
Parameter &DNN& CNN\\\hline
Layer1     &Dense 64-ReLU        & Conv2D 5x8x16-ReLU\\\hline
Layer2     &Dense 128-ReLU       & Conv2D 5x8x32-ReLU\\\hline
Layer3     &Dense 128-ReLU       & Conv2D 5x8x64-ReLU\\\hline
Layer4     &    -                & Dense 1x1x256-ReLU\\\hline
Layer5     &    -                & Dense 1x1x256-ReLU\\\hline
Layer6     &    -                & Dense 1x1x128-ReLU\\\hline
batch size & 128                 & 128\\\hline
epochs     &500                  &100  \\\hline
$\alpha$   & 0.1                 & 0.1\\\hline
$\beta$    & 0.1                 & 0.1\\\hline
\end{tabular}
\label{Training and Testing Parameters}
\end{table}

\subsection{Results Analysis}

\begin{figure}[h]
\centering
\includegraphics[width=3.4in]{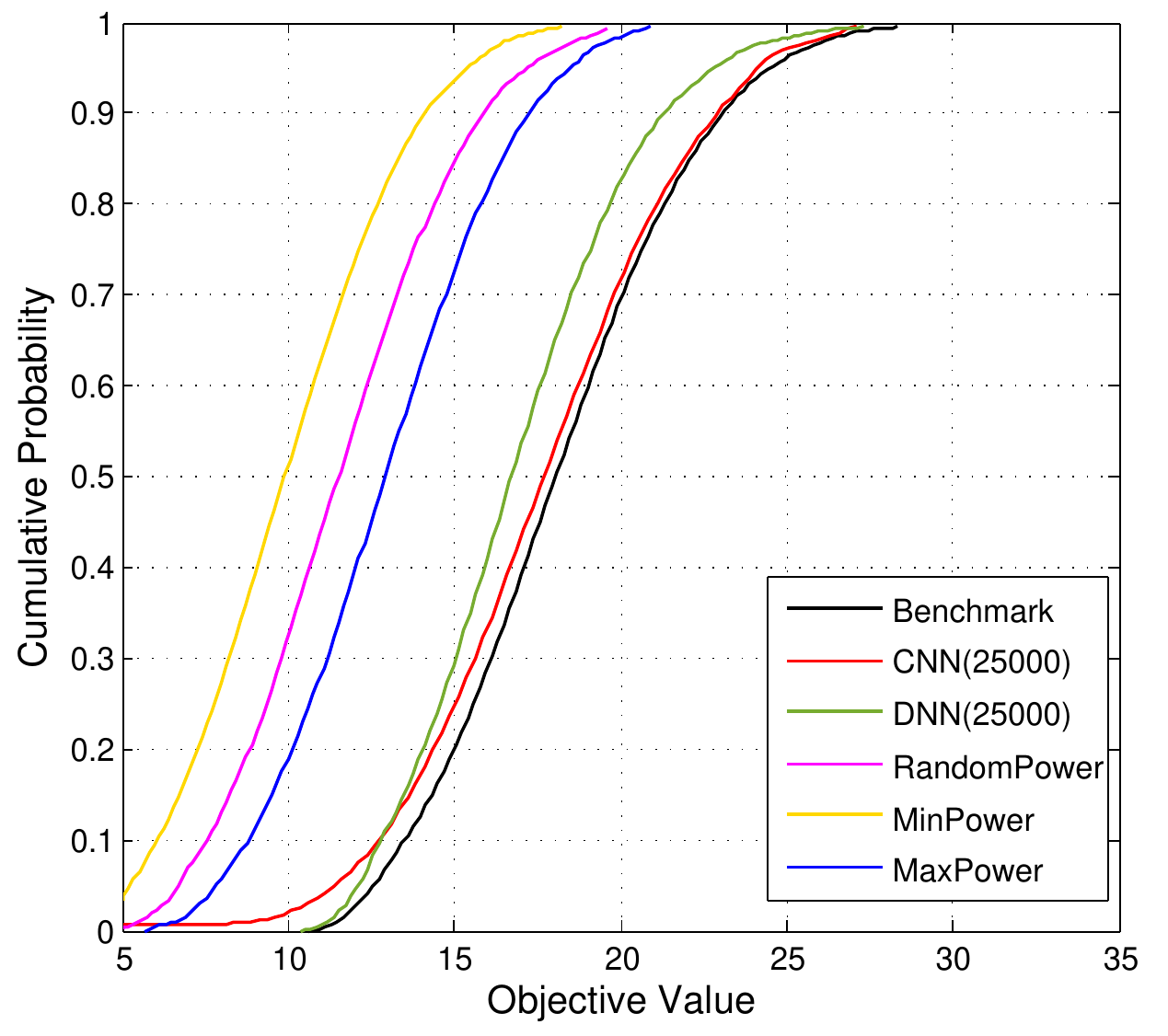}
\caption{The CDF that describes the weighted sum of the ergodic capacity and latency requirement achieved by different approaches: 1) Benchmark; 2) CNN with 25000 training data; 3) DNN with 25000 training data; 4) RandomPower; 5) MinPower; and 6) MaxPower.}
\label{different_power}
\end{figure}

Fig.~\ref{different_power} shows the cumulative distributed function (CDF) of the weighted sum of the ergodic capacity and latency requirement (i.e., the objective in (7)) achieved by different approaches. 
We can see that the proposed CNN approach with 25000 training data outperforms the other schemes with the performance closest to the ideal benchmark method. This is due to the fact that the proposed CNN is expert in extracting the spatial features in channel gains, so as to infer the decisions on resource allocation with a high accuracy.

\begin{figure}[h]
\centering
\includegraphics[width=3.4in]{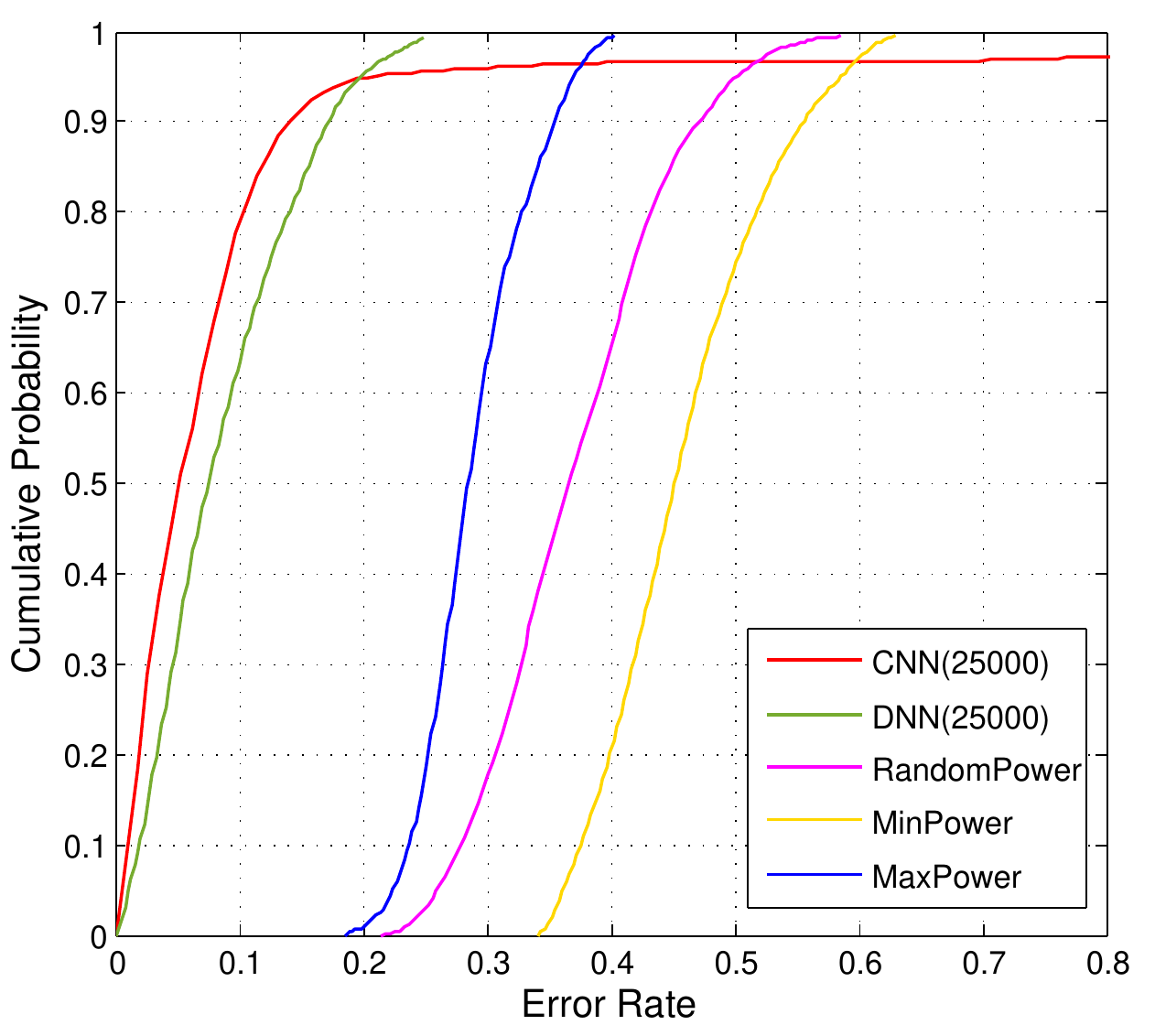}
\caption{The CDF that describes the error rate of different approaches: 1) CNN with 25000 training data; 2) DNN with 25000 training data; 3) RandomPower; 4) MinPower; and 5) MaxPower.}
\label{errorrate_power}
\end{figure}

The superiority of the proposed CNN is further demonstrated by Fig.~\ref{errorrate_power}.
Fig.~\ref{errorrate_power} shows the error rate of different algorithms compared to Benchmark. Let $\eta$ denote the error rate. It can be calculated through 
\begin{eqnarray}
\eta = \frac{C_n - C_0}{C_0},
\end{eqnarray}
where $C_0$ is the objective value in (7) achieved by Benchmark, and $C_n$ is the objective value achieved by the other approaches. We can see that, the error rate of CNN with 25000 training data is smaller than the other approaches. The error rates of 83\% testing data of CNN and 62\% testing data of DNN are within 10\%, while the error rates of the other three algorithms are larger than 20\%. 

\begin{figure}[h]
\centering
\includegraphics[width=3.4in]{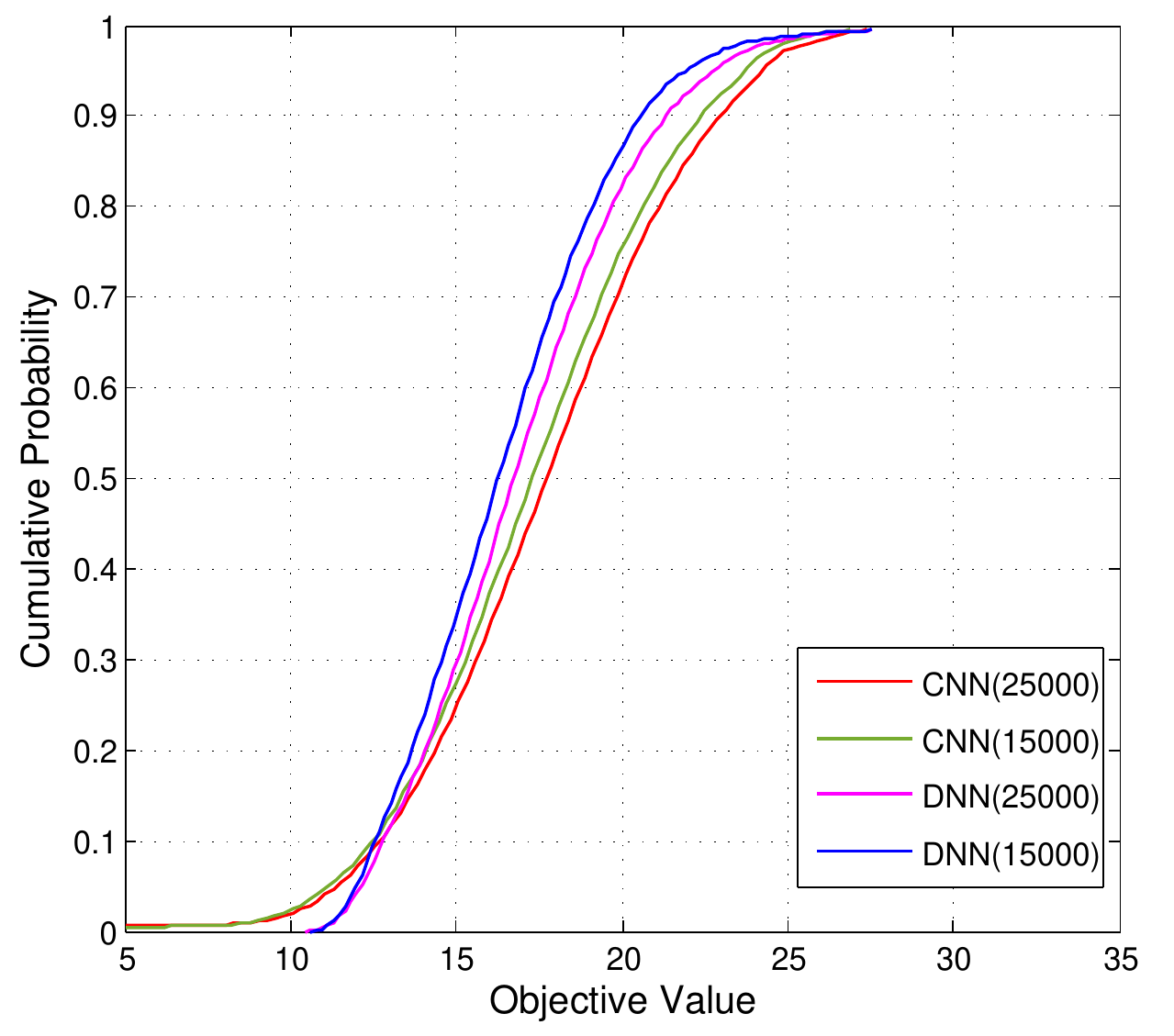}
\caption{The CDF that describes the weighted sum of the ergodic capacity and latency requirement achieved by the neural
networks using different size of training data: 1) CNN with 25000 training data; 2) CNN with 15000 training data; 3) DNN with 25000 training data; and 4) DNN with 15000 training data.}
\label{different_traning}
\end{figure}

\begin{figure}[h]
\centering
\includegraphics[width=3.4in]{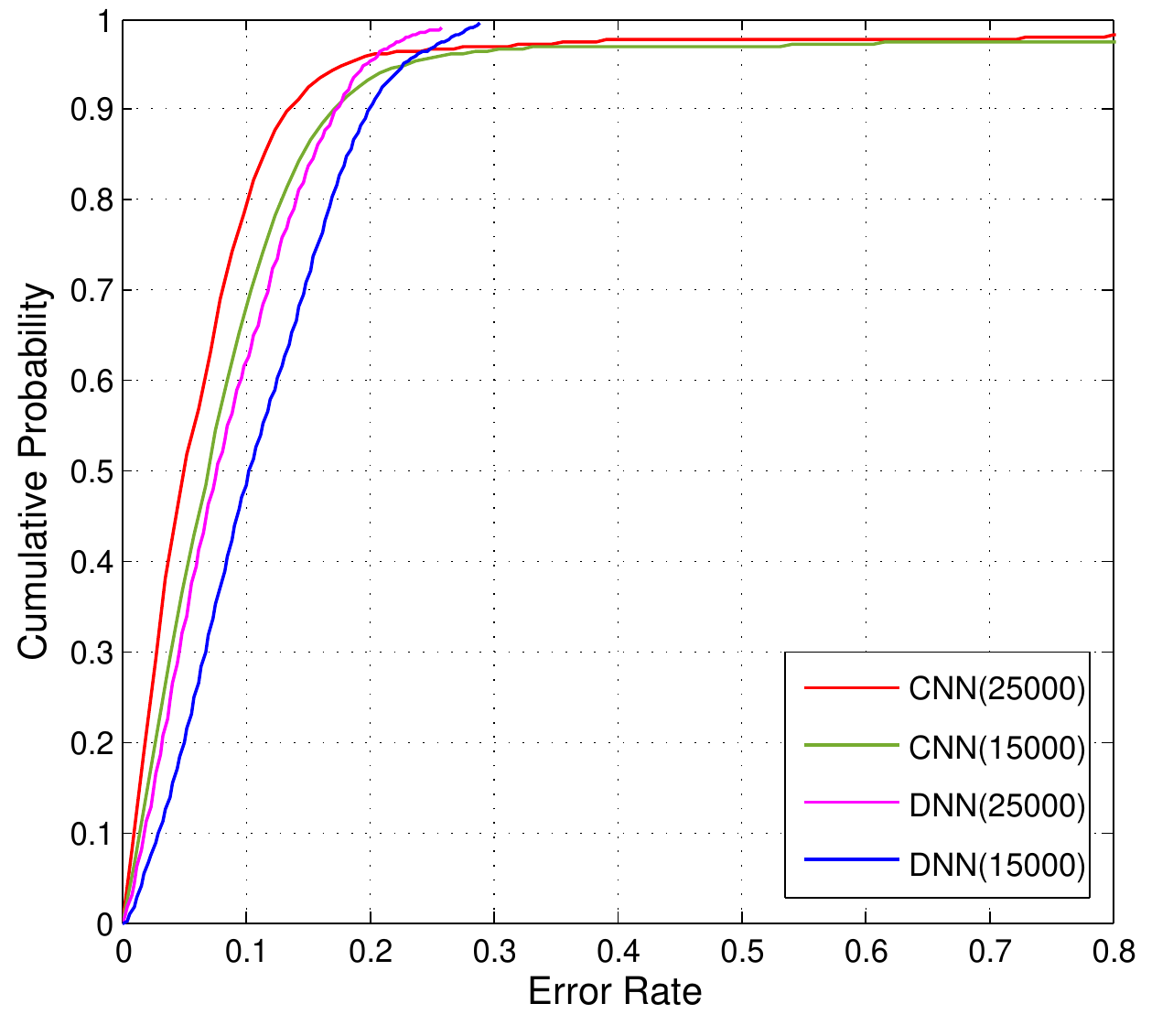}
\caption{The CDF that describes the error rate of the neural
networks using different size of training data: 1) CNN with 25000 training data; 2) CNN with 15000 training data; 3) DNN with 25000 training data; and 4) DNN with 15000 training data.}
\label{ErrorRate}
\end{figure}

Fig.~\ref{different_traning} and Fig.~\ref{ErrorRate} plot the CDF of the objective value and error rate of CNN and DNN with different size of training data, respectively. It can be concluded from the figures that, the CNN-based approach works better than DNN, and a neural network trained with more training data has the performance closer to Benchmark. 

Table~\ref{CPU Runtime} lists the CPU runtime of Benchmark, CNN (with 25000 or 15000 training data), and DNN (with 25000 or 15000 training data). We can see that without compromising much on the performance, CNN with 25000 training data has a CPU runtime only 3.62\% of that with Benchmark. The runtime of CNN is slightly bigger than that of DNN, since we introduce more parameters in CNN for better performance. It is also obvious that more training data results in a larger runtime. This allows us to balance between the runtime and resultant performance.

\begin{table}[h]
\newcommand{\tabincell}[2]{\begin{tabular}{@{}#1@{}}#2\end{tabular}}
\centering
\caption{CPU Runtime Comparison}
\begin{tabular}{|c|c|c|c|c|c|}\hline
Algorithm  &Benchmark   &\tabincell{c}{CNN \\ 25000}    &\tabincell{c}{CNN \\ 15000}  
                        &\tabincell{c}{DNN \\ 25000}    &\tabincell{c}{DNN \\ 15000}  \\\hline
Time (s)     &25.66       &0.93        &0.91      &0.18      &0.17 \\\hline
$\frac{\text{CNN (DNN)}}{\text{Benchmark}}$ & -     & 3.62\%   & 3.54\%   & 0.70\%     & 0.66\% \\\hline
\end{tabular}
\label{CPU Runtime}
\end{table}

\section{Conclusion} \label{sect:conc}
In this paper, we investigated the joint spectrum sharing and power allocation problem for vehicle communication networks that support hybrid V2I and V2V communications. By introducing deep learning techniques, we proposed a CNN-based approach, which decomposed the original problem into a classification subproblem and a regression subproblem, and output the real-time decisions on spectrum reuse and power allocation with a low computational complexity. Extensive numerical experiments demonstrated that the proposed CNN achieved similar performance as the Exhaustive method, while needed only 3.62\% of its CPU runtime.

\section*{Acknowledgment}
This work was supported by the National Natural Science Foundation of China (NSFC) Grants under No. 61701293 and No. 61871262, the National Science and Technology Major Project Grants under No. 2018ZX03001009, the Huawei Innovation Research Program (HIRP), and research funds from Shanghai Institute for Advanced Communication and Data Science (SICS).
\bibliographystyle{IEEEtran}
\bibliography{IEEEfull,ref}
\end{document}